\documentclass[final,3p,times]{elsarticle}

\usepackage{amssymb}
\usepackage{amsmath}
\usepackage{booktabs}%
\usepackage{xcolor}
\usepackage{multirow}
\usepackage{hyperref}
\usepackage{lineno}

\journal{Journal of Neural Engineering}

\begin{document}

\begin{frontmatter}

\title{Spec2VolCAMU-Net: A Spectrogram-to-Volume Model for EEG-to-fMRI Reconstruction based on Multi-directional Time-Frequency Convolutional Attention Encoder and Vision-Mamba U-Net}

\author{Dongyi He}
\author{Shiyang Li}
\author{Bin Jiang\corref{cor1}} 
\author{He Yan}
\cortext[cor1]{Corresponding author. Email: jb20200132@cqut.edu.cn}

\affiliation{organization={School of Artificial Intelligence, Chongqing University of Technology},
            city={Chongqing},
            postcode={400054}, 
            country={China}}

\begin{abstract}
High-resolution functional magnetic resonance imaging (fMRI) is essential for mapping human brain activity; however, it remains costly and logistically challenging. If comparable volumes could be generated directly from widely available scalp electroencephalography (EEG), advanced neuroimaging would become significantly more accessible. Existing EEG-to-fMRI generators rely on plain Convolutional Neural Networks (CNNs) that fail to capture cross-channel time-frequency cues or on heavy transformer/Generative Adversarial Network (GAN) decoders that strain memory and stability. To address these limitations, we propose Spec2VolCAMU-Net, a lightweight architecture featuring a Multi-directional Time-Frequency Convolutional Attention Encoder for rich feature extraction and a Vision-Mamba U-Net decoder that uses linear-time state-space blocks for efficient long-range spatial modelling. We frame the goal of this work as establishing a new state of the art in the spatial fidelity of single-volume reconstruction, a foundational prerequisite for the ultimate aim of generating temporally coherent fMRI time series. Trained end-to-end with a hybrid SSI-MSE loss, Spec2VolCAMU-Net achieves state-of-the-art fidelity on three public benchmarks, recording Structural Similarity Index (SSIM) of 0.693 on NODDI, 0.725 on Oddball and 0.788 on CN-EPFL, representing improvements of 14.5\%, 14.9\%, and 16.9\% respectively over previous best SSIM scores. Furthermore, it achieves competitive Signal-to-Noise Ratio (PSNR) scores, particularly excelling on the CN-EPFL dataset with a 4.6\% improvement over the previous best PSNR, thus striking a better balance in reconstruction quality. The proposed model is lightweight and efficient, making it suitable for real-time applications in clinical and research settings. The code is available at \url{https://github.com/hdy6438/Spec2VolCAMU-Net}.
\end{abstract}


\end{frontmatter}

\section{Introduction}
\label{sec:intro}
The comprehensive delineation of spatiotemporal dynamics in human brain activity is a foundational pursuit in contemporary cognitive and clinical neuroscience, critical for elucidating the neural underpinnings of cognition and a spectrum of neurological and psychiatric disorders \cite{meyer2022spatiotemporal,liu2024activation,belyaeva2024learning}, making non-invasive neuroimaging modalities instrumental in this endeavor. Among these, functional magnetic resonance imaging (fMRI), which infers neural activity via blood-oxygen-level-dependent (BOLD) signal fluctuations, has achieved prominence due to its capacity to generate high-spatial-resolution activation maps \cite{logothetis2008we}. However, the substantial operational demands of fMRI, including cryogenic infrastructure, specialized shielded environments, and expert operating personnel, present significant financial and logistical impediments \cite{constable2023challenges, roos2025brainwaves}. These constraints severely curtail its widespread clinical adoption, limit its utility for real-time neurofeedback applications, and render large-scale, longitudinal investigations economically and practically challenging. Consequently, a compelling imperative exists to develop and validate more accessible and economically viable neuroimaging alternatives that can recapitulate the rich functional information afforded by fMRI.

Electroencephalography (EEG), which directly measures summated postsynaptic electrical potentials manifesting on the scalp, offers exceptional temporal resolution (on the order of milliseconds) and captures cortical neurodynamics with high fidelity \cite{liu2019convolutional}. The inherent advantages of EEG, including its comparatively low cost, portability, and non-invasiveness, have facilitated its extensive deployment across diverse research and clinical paradigms \cite{debener2012taking}. Crucially, a body of evidence from concurrent EEG-fMRI investigations has established robust and neurophysiologically plausible correlations between specific EEG features and BOLD signal dynamics \cite{goldman2002simultaneous, laufs2003eeg, sadaghiani2010intrinsic, chang2013eeg}. These empirical linkages provide a fundamental rationale for exploring the potential to computationally derive high-fidelity, three-dimensional brain activation patterns, analogous to those obtained with fMRI, directly from scalp-recorded EEG signals. Successfully addressing this challenge would represent a significant paradigm shift, potentially democratizing access to advanced functional brain mapping, enabling ubiquitous real-time neural monitoring, and substantially lowering the resource barrier for impactful neuroscientific research globally.

Early pioneering work in this field utilized statistical machine learning methods. This foundational research often fell into two main paradigms: EEG-informed fMRI analysis \cite{cury2020sparse} and joint independent component analysis (ICA) \cite{eichele2009mining}. The former is a hypothesis-driven approach where features derived from EEG, such as event-related potentials or epileptic spikes, are used as regressors in a General Linear Model (GLM) to analyze fMRI data \cite{jamesstatistical}. This allowed researchers to pinpoint fMRI BOLD responses that were spatiotemporally correlated with EEG events. In contrast, joint ICA is a data-driven technique that decomposes both EEG and fMRI data to identify shared, spatially independent components \cite{moeller2011independent}, which helps uncover common brain networks and non-obvious relationships between the two modalities, especially in resting-state studies. These foundational methods paved the way for more targeted predictive models. For example, Meir-Hasson et al. proposed a model based on a time/frequency representation with varying time delays to predict amygdala activity from a single EEG electrode \cite{meir2016one}. This work demonstrated the feasibility of extracting meaningful information from limited EEG data to infer activity in deep brain structures. Similarly, Cury et al. introduced a sparse regression model to predict fMRI-derived neurofeedback scores from EEG signals alone, demonstrating the potential of statistical approaches to bridge the modalities \cite{cury2020sparse}. The use of a sparsity constraint in their model allowed for the selection of the most relevant EEG features, enhancing model interpretability and preventing overfitting. These methods laid the groundwork for leveraging the temporal characteristics of EEG to inform fMRI signals, providing crucial theoretical and methodological precedents for subsequent deep learning-based approaches.

Recent advancements have leveraged deep learning to tackle this complex mapping task, requiring a comprehensive spatiotemporal feature extraction from EEG spectrograms and the synthesis of high-resolution fMRI volumes. Liu and Sajda pioneered the concept of "neural transcoding" using coupled Convolutional Neural Networks (CNNs) to map between EEG and fMRI without explicit physiological priors \cite{liu2019convolutional}, demonstrating feasibility on simulated data, yet their single-scale convolutions cannot capture discriminative cross-channel, cross-frequency patterns, leaving the model vulnerable to noise and acquisition differences, making its lacking robustness to noise and acquisition differences in real scenarios, and the lack of validation on real EEG-fMRI datasets raises concerns about its practical applicability. Later, Calhas et al. investigated autoencoders and Generative Adversarial Networks (GANs), later proposing models incorporating topographical attention mechanisms and Fourier features to explicitly model relationships between EEG electrodes, significantly improving synthesis performance \cite{calhas2022eeg}. However, the random Fourier encoding still entangles temporal and frequency cues, and its multi-branch architecture and random Fourier features lead to a huge hyperparameter space, requiring expensive Bayesian or grid search \cite{zhu2024surface,guo2023adaptive}; More recently, transformer \cite{vaswani2017attention}-based architectures have been introduced, such as the Neural Transcoding Vision Transformer (NT-ViT) \cite{lanzino2024nt} by Lanzino et al., which utilizes a domain-matching module to align latent representations across modalities. However, this kind of Transformer-based  model has a surge in computational complexity when the patch size is reduced or the number of layers is increased \cite{dosovitskiy2020image, pan2022edgevits}, and the author also clearly pointed out that the output can only be used as an "approximation of fMRI" rather than a substitute; Roos et al. proposed E2fNet \cite{roos2025brainwaves}, a U-Net-inspired CNN architecture demonstrating robust performance, particularly in structural similarity, but it still suffers from the limitations of CNNs, such as limited long-range dependencies and high computational costs. Besides, Roos et al. also proposed E2fGAN \cite{roos2025brainwaves}, a GAN-based model \cite{goodfellow2014generative} that generates fMRI volumes from EEG spectrograms. However, GANs are known for their training instability and mode collapse \cite{goodfellow2014generative, salimans2016improved, mescheder2018training}, leading to suboptimal performance in some cases. More recently, some deep learning approaches have explicitly incorporated the time-series nature of the data. For instance, NeuroBOLT, a multi-dimensional transformer-based framework, translates raw EEG time series to fMRI time series by learning representations across temporal, spatial, and spectral domains, enabling more generalized models \cite{li2024neurobolt}. Another work by Li et al. proposed a novel architecture using a Sinusoidal Representation Network (SIREN) to predict fMRI signals directly from multi-channel EEG without explicit feature engineering, focusing on learning frequency information implicitly \cite{li2024leveraging}.

Despite these advancements, existing EEG-to-fMRI synthesis methods face challenges in effective extracting discriminative time-frequency features that capture both local details and cross-channel/global patterns, and modeling long-range spatial dependencies in the fMRI volume without prohibitive memory or training-instability costs. To address the above challenges, we proposed Spec2VolCAMU-Net: First, we propose an EEG encoder named Multi-directional Time-Frequency Convolutional Attention Encoder (MD-TF-CAE) that captures both local and global features from EEG spectrograms. The MD-TF-CAE consists of three parallel convolutional paths: temporal convolution, frequency convolution, and joint time-frequency convolution. Then, the Vision-Mamba U-Net (VM-Unet) \cite{ruan2024vm} decoder is used to replace self-attention with visual state space blocks (VSS Block) \cite{liu2024vmamba}, reducing the complexity of sequence modeling to linear, significantly reducing video memory and inference time, and obtaining smoother gradients and enhanced convergence due to the lack of adversarial targets. With these targeted designs, Spec2VolCAMU-Net simultaneously improves accuracy, stability, and cross-dataset robustness without increasing the complexity of the architecture, providing an efficient and reliable new path for EEG-to-fMRI synthesis, achieving a new state of the art in the spatial fidelity of single-volume reconstruction, which is a foundational prerequisite for the ultimate aim of generating temporally coherent fMRI time series.

Through comprehensive evaluation on multiple publicly available simultaneous EEG-fMRI datasets, including NODDI, Oddball, and CN-EPFL, Spec2VolCAMU-Net demonstrates superior performance compared to existing state-of-the-art approaches. The proposed Spec2VolCAMU-Net consistently achieves the highest Structural Similarity Index (SSIM) across all datasets, indicating exceptional structural fidelity in the reconstructed fMRI volumes. Specifically, Spec2VolCAMU-Net obtained SSIM values of $\textbf{0.693} \pm 0.048$ on NODDI, $\textbf{0.725} \pm 0.040$ on Oddball, and $\textbf{0.788}$ on CN-EPFL, surpassing all compared models. Furthermore, Spec2VolCAMU-Net yielded the highest Signal-to-Noise Ratio (PSNR) on the CN-EPFL dataset ($\textbf{23.837}$), highlighting its effectiveness in minimizing reconstruction errors. These quantitative metrics, complemented by qualitative visualizations, underscore Spec2VolCAMU-Net's ability to generate highly realistic and accurate 3D fMRI volumes from EEG spectrogram inputs.

The contributions of this work are summarized as follows:
\begin{itemize}
  \item We propose Spec2VolCAMU-Net, a novel EEG-to-fMRI synthesis model that effectively captures multi-scale features from EEG spectrograms using a Multi-directional Time-Frequency Convolutional Attention Encoder (MD-TF-CAE) and reconstructs fMRI volumes using a Vision-Mamba U-Net decoder.
  \item The proposed MD-TF-CAE captures both local and global features from EEG spectrograms through multi-directional convolutions and multi-head self-attention, addressing the challenges of comprehensive feature extraction and long-range dependency modeling.
  \item We demonstrate superior performance of Spec2VolCAMU-Net compared to existing state-of-the-art methods across multiple public EEG-fMRI datasets (NODDI, Oddball, and CN-EPFL), particularly in achieving higher SSIM and competitive PSNR, indicating improved structural fidelity and detail accuracy in reconstructed fMRI volumes.
\end{itemize}

The remainder of this paper is organized as follows: Section \ref{sec:model} presents the proposed Spec2VolCAMU-Net architecture and its components. Section \ref{sec:exp} describes the experimental setup, including data preprocessing, datasets, baseline methods, evaluation metrics, and implementation details. Section \ref{sec:result} presents the experimental results and discusses the performance of Spec2VolCAMU-Net compared to baseline methods. Finally, Section \ref{sec:conclusion} concludes the paper and outlines future work.

\section{Method}
\label{sec:model}

\subsection{Problem Definition}
\label{sec:problem}
Given a set of EEG spectrograms $x \in \mathbb{R}^{C \times F \times T}$, where $C$ is the number of channels, $F$ is the number of frequency bands, and $T$ is the number of time steps, the goal is to learn a mapping function $\mathcal{F}(x)$ that generates a corresponding fMRI volume $y \in \mathbb{R}^{D \times H \times W}$, where $D$, $H$, and $W$ are the depth of the fMRI volume, height, and width, respectively.

\subsection{The architecture of Spec2VolCAMU-Net}
\label{sec:architecture}
\begin{figure*}[t]
  \centering
  \includegraphics[width=0.9\textwidth]{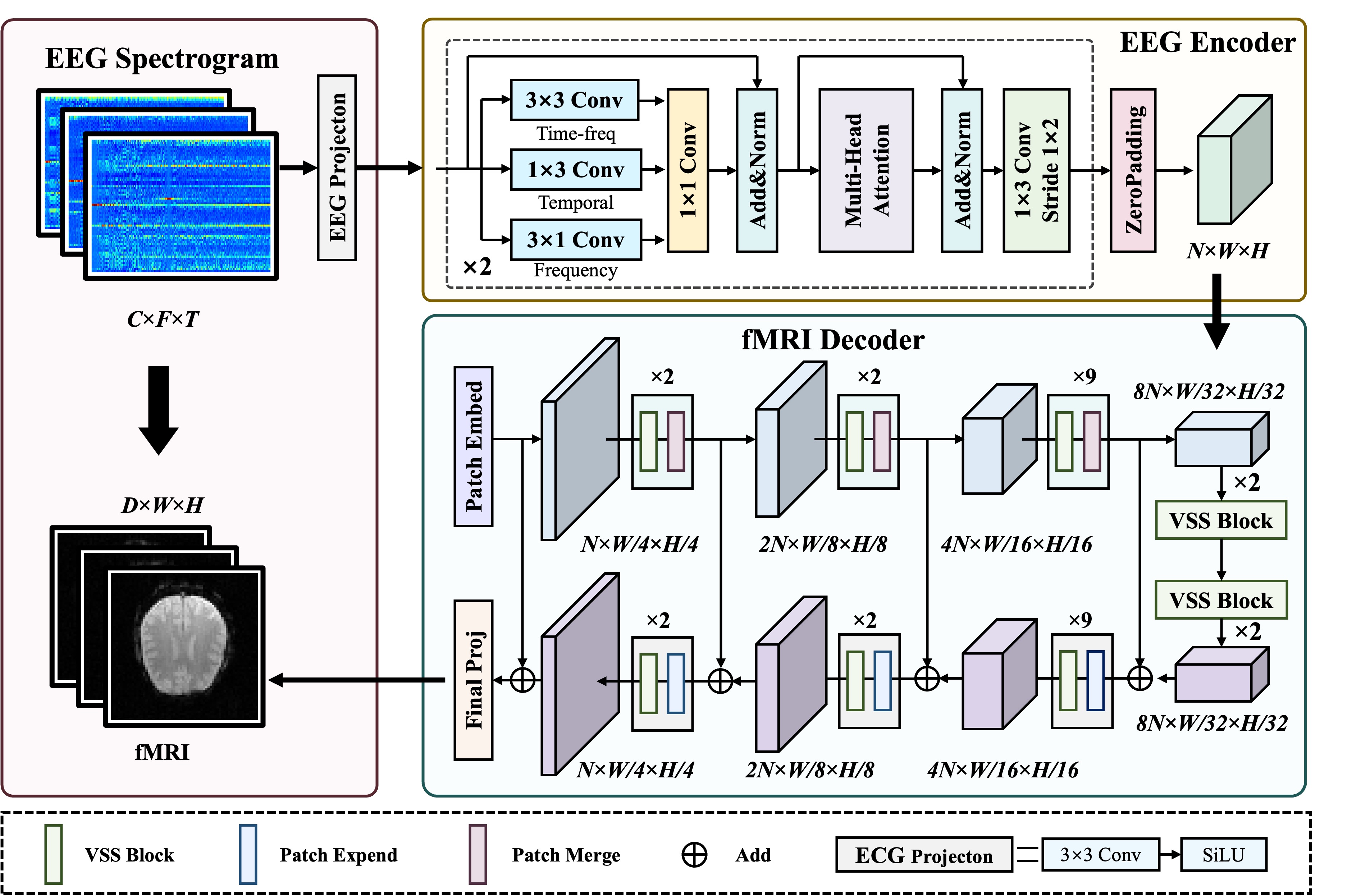}
  \caption{The architecture of Spec2VolCAMU-Net, which consists of a proposed Multi-directional Time-Frequency Convolutional Attention Encoder (MD-TF-CAE) and a Vision-Mamba U-Net decoder.}
  \label{fig:model}
\end{figure*}
The architecture of Spec2VolCAMU-Net is shown in Figure \ref{fig:model}. The model consists of two main components: a Multi-directional Time-Frequency Convolutional Attention Encoder (MD-TF-CAE) and a Vision-Mamba U-Net decoder. The encoder captures local features from the projection of EEG spectrograms using lightweight convolutional layers, followed by multi-head self-attention to integrate global correlations across electrodes. This design alleviates the generalization and parameter redundancy issues associated with small sample scenarios. The decoder employs a Vision-Mamba U-Net architecture, which replaces self-attention with selective state-space blocks. This modification reduces the complexity of sequence modeling to linear, significantly decreasing video memory usage and inference time. Additionally, the absence of adversarial targets leads to smoother gradients and enhanced convergence stability.

\subsubsection{EEG Spectrogram Projection}
\label{sec:projection}
The EEG spectrogram projection module maps the input EEG tensor $x \in \mathbb{R}^{C \times T \times F}$ to a high-dimensional feature space using a 3x3 convolutional kernel and SiLU activation function, resulting in an initial embedding representation $x' \in \mathbb{R}^{N \times T \times F}$. This processing aims to enhance the nonlinear expressiveness of the signal and provide a foundation for subsequent feature extraction.

\subsubsection{Multi-directional Time-Frequency Convolutional Attention Encoder (MD-TF-CAE)}
\label{sec:encoder}
In the EEG encoder module, the proposed MD-TF-CAE, Multi-directional Time-Frequency Convolutional Attention Encoder, first performs local feature extraction: $x'$ is input into three parallel convolutional paths, including temporal convolution ($3\times1$ convolutional layer), frequency convolution ($1\times3$ convolutional layer), and joint time-frequency convolution ($3\times3$ convolutional layer), capturing local structural information of the EEG signal in different dimensions. The multi-path features are then concatenated along the channel dimension and fused through a point-wise convolution. The fused features are added to the original input $x'$ to construct a residual connection structure that enhances gradient propagation. Layer normalization is applied to stabilize the model training process, resulting in a local feature representation $F_{local}$.

Then, the encoder enters the global feature extraction stage. The local feature representation $F_{local}$ is fed into a multi-head self-attention module (MHSA) to capture long-range dependencies across time and frequency dimensions. The output of the MHSA is again connected to $F_{local}$ through a residual connection, followed by layer normalization, forming a feature representation with global perceptual capabilities. Subsequently, a frequency-direction convolutional layer (with a $1\times3$ kernel and a stride of 2) is applied for downsampling in the frequency dimension. This EEG encoding process is repeated twice, ultimately yielding a multi-level fused global feature representation $F_{global}$. To align with the target fMRI image space, zero-padding is used to adjust $F_{global}$ to the target size, resulting in the feature map $F \in \mathbb{R}^{N \times H \times W}$.

\subsubsection{Vision-Mamba U-Net Decoder}
\label{sec:decoder}
For the final synthesis of the fMRI volume, the aligned feature map $F$ is processed by a Vision-Mamba U-Net (VM-UNet) \cite{ruan2024vm} decoder. This decoder is responsible for multi-scale feature decoding and spatial reconstruction, ultimately transforming the $N$-channel feature map into the final three-dimensional fMRI volume $Y \in \mathbb{R}^{D \times H \times W}$. The architecture of the VM-UNet is designed to synergize the capacity for local feature modeling, typically associated with CNNs, with the efficient long-range dependency modeling of Mamba modules, making it highly effective for detailed image reconstruction.

Instead of relying on traditional convolutions, the VM-UNet decoder is built upon Visual State Space (VSS) blocks \cite{liu2024vmamba}. At the core of each VSS block is the 2D-Selective-Scan (SS2D) module, a Mamba variant adapted for vision tasks \cite{ruan2024vm, gu2023mamba}. The SS2D mechanism is engineered to capture spatial information efficiently. It operates by first expanding the input feature map into four separate sequences by scanning it in four different directions (e.g., top-left to bottom-right). Each sequence is then processed by a structured state-space model (S6 block) \cite{gu2023mamba} to extract features. Finally, a merging step combines the information from all four directional scans, producing an output feature map of the same dimensions as the input but enriched with multi-directional context.

\subsection{Loss Function}
\label{sec:loss}
The Spec2VolCAMU-Net model is trained using a supervised end-to-end optimization strategy, aiming to construct a high-quality mapping model from EEG signals to fMRI images. To enhance the structural restoration capability and pixel-level reconstruction accuracy of the synthesized images, a joint loss function is designed, combining Structural Similarity Loss (SSIM Loss) and Mean Squared Error Loss (MSE Loss), defined as follows:
\begin{equation}
  \mathcal{L}_{total} = \lambda_{SSIM} \cdot \mathcal{L}_{SSIM} + \lambda_{MSE} \cdot \mathcal{L}_{MSE}
\end{equation}
where $\lambda_{SSIM},\lambda_{MSE}\in\mathbb{R}_{\ge 0}$ are weighting coefficients, and $\lambda_{SSIM} + \lambda_{MSE} = 1$.

The SSIM loss is defined as $\mathcal{L}_{\text{SSIM}} = 1 - \text{SSIM}(x,y)$ so that minimizing $\mathcal{L}_{\text{SSIM}}$ directly maximizes structural affinity to preserve neuro-anatomical structure, whereas $\mathcal{L}_{\text{MSE}}$ penalizes voxel-wise intensity deviations, sharpening contrast and fine-grained details. By jointly optimizing $\mathcal{L}_{\text{SSIM}}$ and $\mathcal{L}_{\text{MSE}}$, provides complementary gradients: SSIM aligns the macro-scale spatial layout, whereas MSE fine-tunes micro-scale pixel accuracy. This synergy can be encouraged to preserve salient neuroanatomical structures while faithfully reconstructing fine-grained details, yielding reconstructions that are both perceptually and quantitatively superior.

\section{Experimental Setting}
\label{sec:exp}
\subsection{Data Perprocessing}
\label{sec:preprocessing}
To efficiently train the proposed Spec2VolCAMU-Net, it is essential to construct a set of structurally consistent and temporally aligned multimodal input-output samples. This paper is based on paired EEG-fMRI open-access dataset and designs a systematic data preprocessing pipeline to ensure that the extracted EEG time-frequency features are aligned with the fMRI images in the temporal dimension and possess learning value in the feature space. Specifically, each channel's EEG signal is divided into window sequences of length $fs \times TR$, where $fs$ is the sampling rate and $TR$ is the time resolution of fMRI, to achieve correspondence between EEG and fMRI sampling periods.

Subsequently, a Short-Time Fourier Transform (STFT) is applied to each EEG window to extract its time-frequency representation, obtaining a more structured time-frequency spectrogram covering the 20 s window that ends 6 s before each BOLD slice, compensating for neuro-vascular lag. To remove high-frequency noise and enhance signal stability, the STFT results are filtered using a bandpass filter with a cutoff frequency of 250 Hz. Additionally, to eliminate the influence of baseline drift on model training, the direct current (0 Hz) component is further removed. Finally, the EEG input samples are constructed as a three-dimensional tensor $X \in \mathbb{R}^{C \times T \times F}$, where $C$ is the number of electrode channels, $T$ is the number of time windows, and $F$ is the frequency dimension. This tensor is paired with the corresponding fMRI three-dimensional volume image $Y \in \mathbb{R}^{D \times H \times W}$ as supervised signal input. To unify the numerical scale of each modality and improve model convergence efficiency, all EEG and fMRI data are processed using max-min normalization, mapping them to the range $\left[0,1\right]$.

\subsection{Dataset}
\label{sec:dataset}
In this study, we evaluate the performance of Spec2VolCAMU-Net on three publicly available EEG-fMRI datasets: NODDI \cite{lanzino2024nt}, Oddball \cite{calhas2022eeg}, and CN-EPFL \cite{roos2025brainwaves}. These datasets are selected for their diverse characteristics and have been widely used in previous research on EEG-to-fMRI synthesis. Table \ref{tab:dataset} summarizes the key characteristics of each dataset, including the number of participants, EEG and fMRI acquisition parameters, and the number of paired samples.

The NODDI dataset is a resting-state EEG-fMRI dataset contains simultaneous recordings from 17 healthy adults who fixated on a white cross for about 11 min. EEG was acquired with a 64-channel 10-20 cap at 250 Hz, while whole-brain fMRI volumes were collected on a Siemens Avanto 1.5 T scanner ($TR = 2.16 s$, \(30\times64\times64\) voxels, \(3{\times}3{\times}3\) mm). Following common practice, two subjects with excessive motion were discarded, leaving 15 participants and about 4,110 paired EEG-fMRI samples after temporal alignment and windowing.

The Oddball dataset provides target-detection recordings from the same number of subjects (17), but during rapid auditory and visual oddball trials. EEG was sampled at 1 kHz from 43 scalp electrodes; fMRI was acquired on a Philips Achieva 3 T with TR = 2 s and \(32\times64\times64\) voxels. After synchronising each 2 s EEG window to the subsequent BOLD volume we obtain 1 020 volumes per subject, totalling 17 340 paired examples.

The CN-EPFL consists of 20 subjects performing a speeded visual discrimination task with trial-by-trial confidence ratings. EEG was recorded at 5 kHz from 64 channels, and fMRI volumes were captured on a Siemens Trio 3 T (TR = 1.28 s, \(54\times108\times108\) voxels, \(2{\times}2{\times}2\) mm). To harmonise spatial resolution with the other benchmarks we follow earlier work and apply a DCT-based down-sampling to \(30\times64\times64\), yielding 6 880 EEG-fMRI pairs.

The diversity in paradigm, field strength (1.5 T vs.\ 3 T) and sampling rates (250 Hz-5 kHz) ensures that improvements observed with Spec2VolCAMU-Net cannot be attributed to dataset-specific quirks alone.

\begin{table*}[h]
\centering
\caption{Key characteristics of the three evaluation datasets.}
\begin{tabular}{lcccc}
\toprule
Dataset & Participants & EEG (ch./Hz) & fMRI (TR / voxels) & Pairs \\
\midrule
NODDI & 15 & 64 / 250 & 2.16 s / \(30{\times}64{\times}64\) & 4 110 \\
Oddball & 17 & 43 / 1 000 & 2 s / \(32{\times}64{\times}64\) & 17 340 \\
CN-EPFL & 20 & 64 / 5 000 & 1.28 s / \(54{\times}108{\times}108\) & 6 880 \\
\bottomrule
\end{tabular}
 \label{tab:dataset}
\end{table*}

\subsection{Baseline Methods}
\label{sec:baseline}
We compare the performance of Spec2VolCAMU-Net with several state-of-the-art EEG-to-fMRI synthesis methods, including:
\begin{itemize}
  \item \textbf{CNN-TC}~\cite{liu2019convolutional}.  
  The first neural transcoding model employs a pair of symmetric CNNs in which the EEG\,$\rightarrow$\,fMRI branch stacks 1-D temporal kernels (length~27) with 3-D spatial kernels to regress a full \(D{\times}W{\times}H\) volume from 20 s of EEG. Despite being trained only on simulation in the original paper, the architecture remains a strong low-parameter benchmark when ported to real data.
  
  \item \textbf{CNN-TAG}~\cite{calhas2022eeg}.  
  Calhas \emph{et al.} augment a multi-branch CNN encoder with (i) a topographical attention graph that attends over electrode-electrode relations, and (ii) random Fourier feature projections whose scale is modulated  by the attention scores. These additions help the network capture long-range scalp interactions that map onto distributed BOLD patterns.
  
  \item \textbf{NT-ViT}~\cite{lanzino2024nt}.  
  A Vision-Transformer generator that tokenises 3-D fMRI patches and EEG Mel-spectrogram patches, then enforces domain matching between their latent tokens. On NODDI and Oddball the authors report a \(10\times\) lower RMSE and \(3.1\times\) higher SSIM than earlier CNNs, setting a new state-of-the-art for transformer baselines.
  
  \item \textbf{E2fGAN}~\cite{roos2025brainwaves}.  
  A conditional GAN that re-uses the E2fNet generator (below) and adds a 3-D PatchGAN discriminator. Although adversarial training can yield locally sharper volumes, Roos et al. note frequent convergence instabilities that degrade mean SSIM/PSNR, especially on the small NODDI split.
  
  \item \textbf{E2fNet}~\cite{roos2025brainwaves}.  
  A lightweight encoder-U-Net-decoder crafted for EEG spectrograms: the encoder expands the temporal axis to 256 feature maps while preserving electrode topology; a two-down / two-up U-Net refines multi-scale information before a decoder squeezes depth back to \(D\). Trained with mixed SSIM+MSE loss, E2fNet currently delivers the best mean SSIM across all three public datasets (e.g.\ 0.631 on Oddball vs.\ 0.189 for CNN-TC).
\end{itemize}

These models represent a range of approaches, including CNNs, transformers, and GANs, providing a comprehensive comparison to evaluate the effectiveness of Spec2VolCAMU-Net.

\subsection{Evaluation Metrics}
\label{sec:metrics}
To quantitatively evaluate the performance of Spec2VolCAMU-Net and the baseline methods, we employ two widely used metrics in image synthesis tasks: Structural Similarity Index (SSIM) and Peak Signal-to-Noise Ratio (PSNR). SSIM measures the structural similarity between the synthesized fMRI volume and the ground truth, while PSNR quantifies the peak error between them. These metrics provide insights into both the perceptual quality and fidelity of the generated fMRI volumes.
\begin{equation}
  \text{SSIM}(x,y) = \frac{(2\mu_x\mu_y + c_1)(2\sigma_{xy} + c_2)}{(\mu_x^2 + \mu_y^2 + c_1)(\sigma_x^2 + \sigma_y^2 + c_2)}
\end{equation}
where \(x\) and \(y\) are the synthesized and ground truth images, respectively; \(\mu_x\) and \(\mu_y\) are the mean intensities of \(x\) and \(y\); \(\sigma_x^2\) and \(\sigma_y^2\) are the variances of \(x\) and \(y\); \(\sigma_{xy}\) is the covariance between \(x\) and \(y\); and \(c_1\) and \(c_2\) are small constants to avoid division by zero.
\begin{equation}
  \text{PSNR}(x,y) = 10 \cdot \log_{10}\left(\frac{\text{MAX}^2}{\text{MSE}(x,y)}\right)
\end{equation}
where \(\text{MAX}\) is the maximum possible pixel value of the image, which is 1 for normalized images, and \(\text{MSE}(x,y) = \frac{1}{N}\sum_{i=1}^{N}(x_i - y_i)^2\) is the mean squared error between \(x\) and \(y\), and \(N\) is the number of pixels in the image.

\subsection{Implementation Details}
\label{sec:implementation}
The Spec2VolCAMU-Net model is implemented using PyTorch and trained on a single NVIDIA GeForce RTX 4070 GPU. The model is trained for 50 epochs with a batch size of 16, using the AdamW optimizer with an initial learning rate of \(1 \times 10^{-3}\) and a weight decay of \(1 \times 10^{-2}\). A cosine annealing schedule with hard restarts is applied, where the learning rate restarts every 10 epochs. The loss function is a combination of SSIM and MSE losses, weighted by \(\lambda_{SSIM}\) and \(\lambda_{MSE}\).

\begin{figure}[h]
  \centering
  \includegraphics[width=0.8\textwidth]{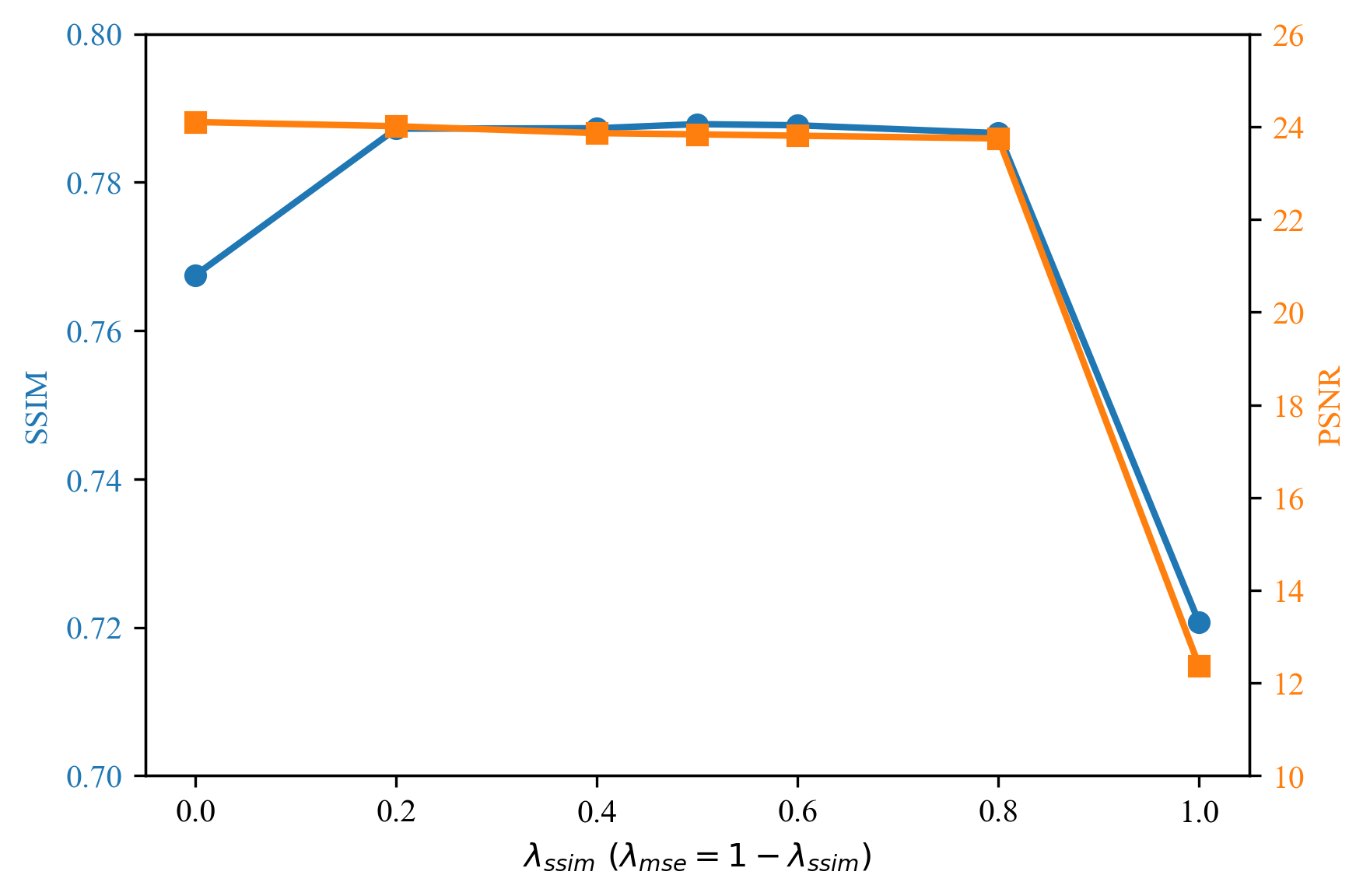}
  \caption{Trade-off between SSIM and PSNR on the validation set for different values of \(\lambda_{SSIM}\) and \(\lambda_{MSE}\). The optimal balance is achieved at \(\lambda_{SSIM} = 0.5\) and \(\lambda_{MSE} = 0.5\).}
  \label{fig:tradeoff}
\end{figure}

To determine the optimal values of \(\lambda_{SSIM}\) and \(\lambda_{MSE}\), a grid search was performed over the range \([0, 1]\) with a step size of 0.2, ensuring that \(\lambda_{SSIM} + \lambda_{MSE} = 1\). The model was evaluated on a CN-EPFL dataset using SSIM and PSNR as metrics, and the combination of \(\lambda_{SSIM} = 0.5\) and \(\lambda_{MSE} = 0.5\) yielded the best overall performance across all datasets. The trade-off between SSIM and PSNR for different values of \(\lambda_{SSIM}\) and \(\lambda_{MSE}\) is illustrated in Figure \ref{fig:tradeoff}.

Mirroring previous studies, the model is evaluated using leave-one-subject-out (LOSO) cross-validation on the NODDI and Oddball datasets, where one subject is held out for testing while the remaining subjects are used for training. For the CN-EPFL dataset, we follow the conventional 16/4 train-test split, where 16 subjects are used for training and 4 subjects are used for testing. The performance of each model is reported as the mean and standard deviation of SSIM and PSNR across all test subjects.

\subsection{Ablation Study}
\label{sec:ablation}
To validate the effectiveness of the proposed components in Spec2VolCAMU-Net, we conduct an ablation study on the CN-EPFL dataset. We systematically remove or replace key components of the model and evaluate the impact on performance using SSIM and PSNR metrics. Specifically, we replace the Vision-Mamba U-Net decoder with a standard U-Net architecture, and remove each of the three convolutional paths in the MD-TF-CAE (temporal, frequency, and joint time-frequency), as well as the multi-head self-attention module.

\section{Results and Discussion}
\label{sec:result}
\subsection{\textcolor{blue}{Performance Comparison with State-of-the-Art Methods}}
\begin{table*}[htbp]
  \centering
  \caption{Performance comparison across datasets using SSIM and PSNR.}
  \begin{tabular}{lcccccc}
    \toprule
   \multirow{2}{*}{\textbf{Model}} & \multicolumn{2}{c}{\textbf{NODDI}} & \multicolumn{2}{c}{\textbf{Oddball}} & \multicolumn{2}{c}{\textbf{CN-EPFL}} \\
    \cmidrule(lr){2-3} \cmidrule(lr){4-5} \cmidrule(lr){6-7}  
   & SSIM & PSNR & SSIM & PSNR & SSIM & PSNR \\
   \midrule
  CNN-TC \cite{liu2019convolutional} & 0.449 $\pm$ 0.060 & --- & 0.189 $\pm$ 0.038 & --- & 0.519 & --- \\
  CNN-TAG \cite{calhas2022eeg} & 0.472 $\pm$ 0.010 & --- & 0.200 $\pm$ 0.017 & --- & 0.522 & --- \\
  NT-ViT \cite{lanzino2024nt}  & 0.581 $\pm$ 0.048 & \textcolor{red}{\textbf{21.56 $\pm$ 1.06}} & 0.627 $\pm$ 0.051 & \textcolor{red}{\textbf{23.33 $\pm$ 1.04}} & --- & --- \\
E2fGAN \cite{roos2025brainwaves} & 0.576 $\pm$ 0.047 & 18.535 $\pm$ 1.345 & 0.583 $\pm$ 0.034 & 18.711 $\pm$ 1.754 & 0.607 & 18.172 \\
E2fNet \cite{roos2025brainwaves}  & \textcolor{blue}{\textbf{0.605 $\pm$ 0.046}} & 20.096 $\pm$ 1.280 & \textcolor{blue}{\textbf{0.631 $\pm$ 0.042}} & 22.193 $\pm$ 1.013 & \textcolor{blue}{\textbf{0.674}} & \textcolor{blue}{\textbf{22.781}} \\
Spec2VolCAMU-Net & \textcolor{red}{\textbf{0.693 $\pm$ 0.048}} & \textcolor{blue}{\textbf{20.803 $\pm$ 1.143}} & \textcolor{red}{\textbf{0.725 $\pm$ 0.040}} & \textcolor{blue}{\textbf{22.737 $\pm$ 0.968}} & \textcolor{red}{\textbf{0.788}}&	\textcolor{red}{\textbf{23.837}}\\
\toprule
\end{tabular}
  \label{tab:comparison}
  \footnotesize{
    The best results are highlighted in \textcolor{red}{\textbf{red}} and the second-best results are highlighted in \textcolor{blue}{\textbf{blue}}.\\
    $^*$ The results for other methods are taken from their respective papers.
    }
\end{table*}

The proposed Spec2VolCAMU-Net model demonstrates a significant advancement in the field of EEG-to-fMRI reconstruction, consistently achieving state-of-the-art performance in structural similarity across multiple challenging public datasets. As detailed in Table~\ref{tab:comparison}, Spec2VolCAMU-Net markedly surpasses existing leading methods, including E2fNet, NT-ViT, and E2fGAN, in terms of the SSIM on the NODDI, Oddball, and CN-EPFL datasets. Specifically, the model achieved SSIM scores of $0.693 \pm 0.048$ on NODDI, $0.725 \pm 0.040$ on Oddball, and $0.788$ on CN-EPFL, representing improvements of 14.5\% ($p\ll0.001$, 95\% CI of difference [0.053,0.122]), 14.9\% ($p\ll0.001$, 95\% CI of difference [0.065,0.123]), and 16.9\% respectively, over the previous best SSIM scores reported by E2fNet. This substantial enhancement in SSIM is critical as it indicates a superior capability to preserve the neuro-anatomical structure and spatial layout of the brain in the synthesized fMRI volumes, a key objective in cross-modal neuroimaging translation.

\begin{figure*}[t]
  \centering
  \includegraphics[width=0.9\textwidth]{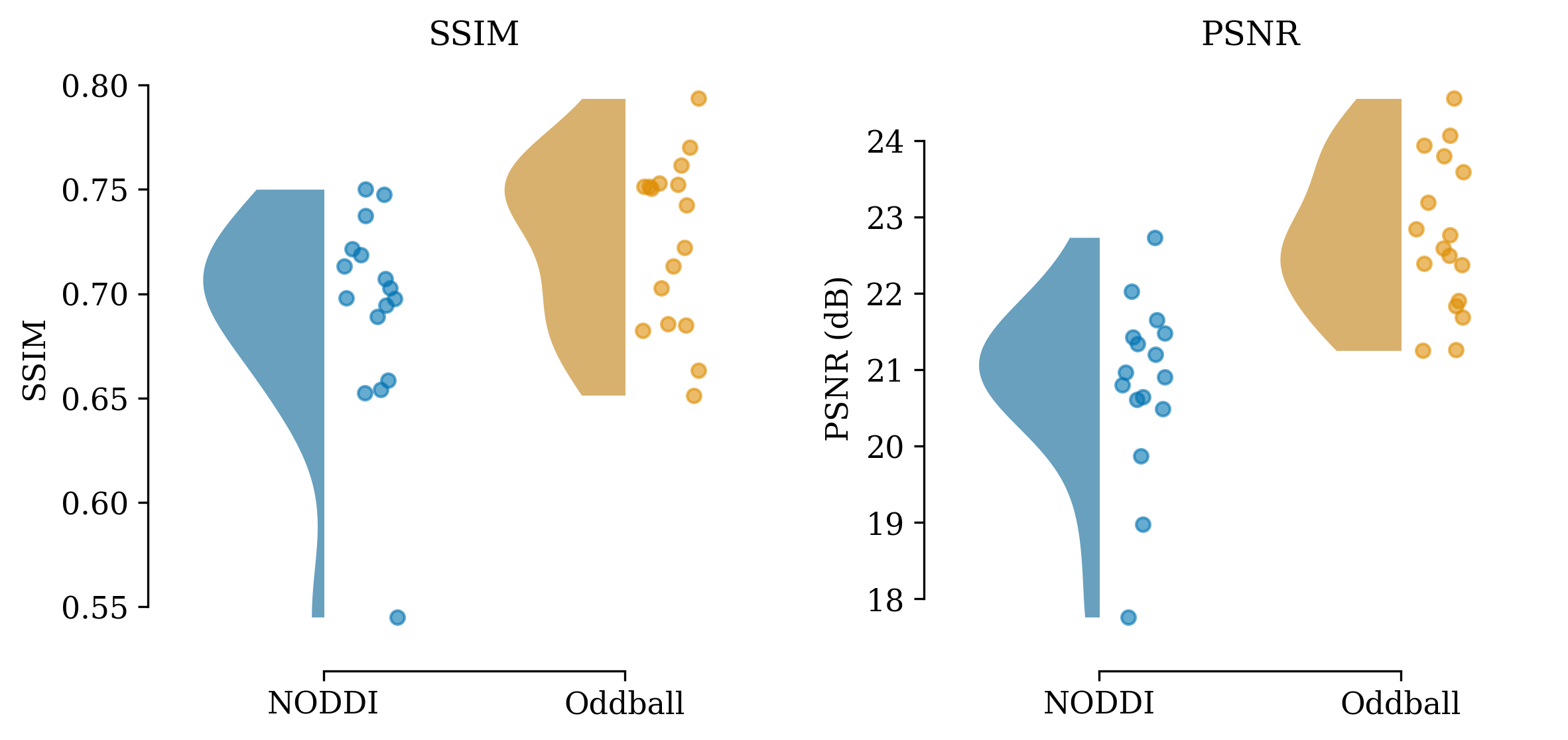}
  \caption{Half-violin plots (left half ) with overlad jittered scatter points (right half) summarise the distribution of SSIM and PSNR scores for the proposed Spec2VolCAMU-Net accross NODDI and Oddball datasets.}
  \label{fig:half_violin}
\end{figure*}

In addition to its leading SSIM performance, Spec2VolCAMU-Net achieves highly competitive PSNR scores, particularly excelling on the CN-EPFL dataset with a PSNR of 23.837, which is 4.6\% higher than E2fNet, indicating improved detail fidelity and reduced voxel-wise intensity deviations. While NT-ViT reported marginally higher PSNR values on the NODDI (21.56 dB) and Oddball (23.33 dB) datasets, this came at the cost of considerably lower SSIM scores (0.581 and 0.627, respectively). This comparison underscores a crucial advantage of Spec2VolCAMU-Net: it strikes a more effective balance between preserving structural integrity, SSIM, and ensuring voxel-level accuracy, PSNR. The ability to generate reconstructions that are both structurally coherent and rich in detail is paramount for the practical utility of synthesized fMRI data in neuroscientific research and potential clinical applications. The half-violin plots in Figure~\ref{fig:half_violin} further illustrate the robustness of Spec2VolCAMU-Net, showing tightly clustered SSIM and PSNR distributions, particularly on the Oddball dataset, and commendable performance even with lower field strength and noisier signals from the NODDI dataset.

\subsection{Qualitative Visual Comparisons}
\begin{figure*}[t]
  \centering
  \includegraphics[width=0.8\textwidth]{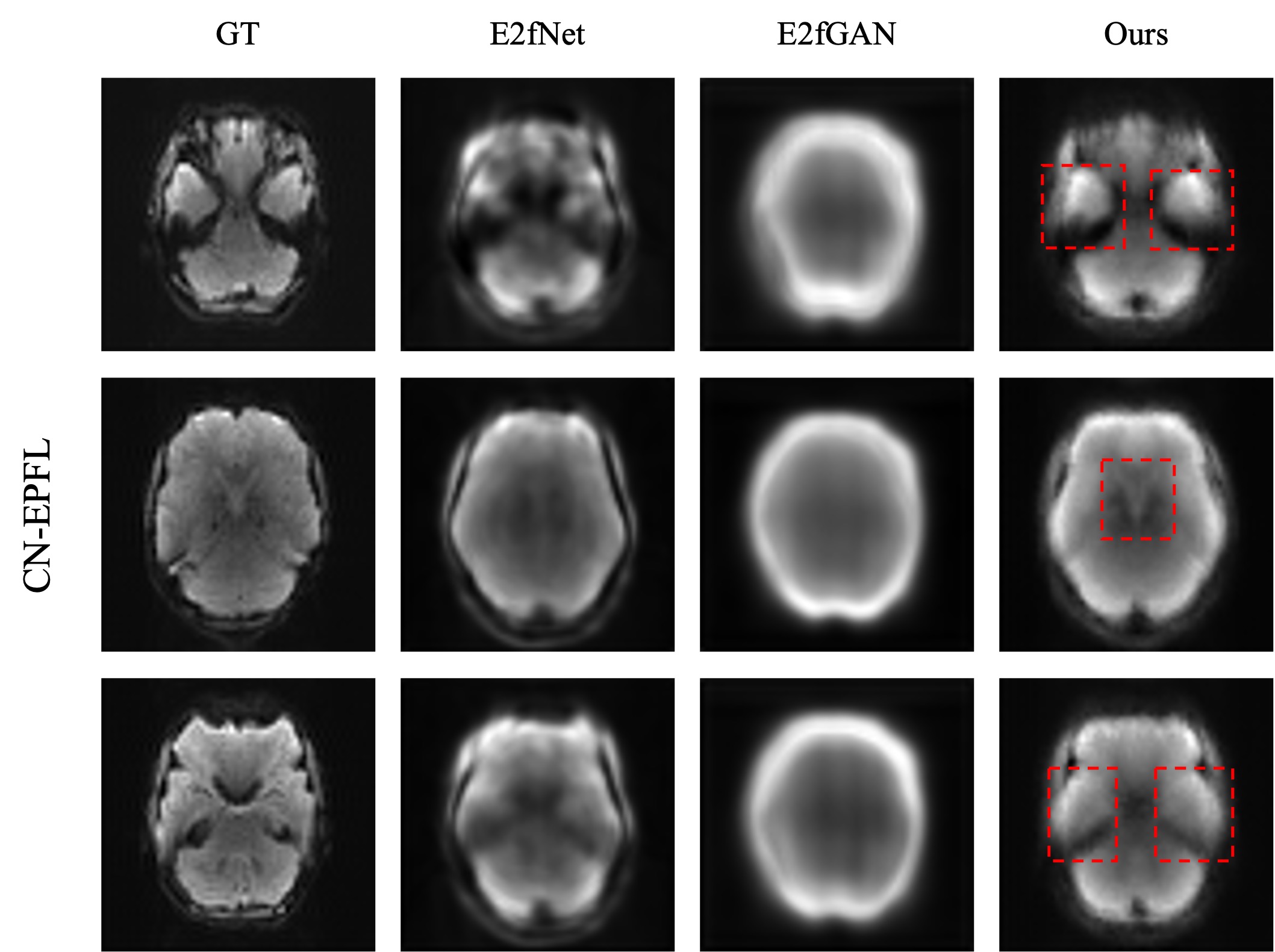}
  \caption{Visual comparison of reconstructed fMRI volumes from Spec2VolCAMU-Net and certain baseline methods on the CN-EPFL dataset. The first column shows the ground truth (GT) fMRI volume, while the subsequent columns display the results from E2fNet, E2fGAN, and Spec2VolCAMU-Net. The brain regions that improve the most with Spec2VolCAMU-Net are highlighted with red boxes.}
  \label{fig:cn-epfl}
\end{figure*}

\begin{figure*}[t]
  \centering
  \includegraphics[width=0.8\textwidth]{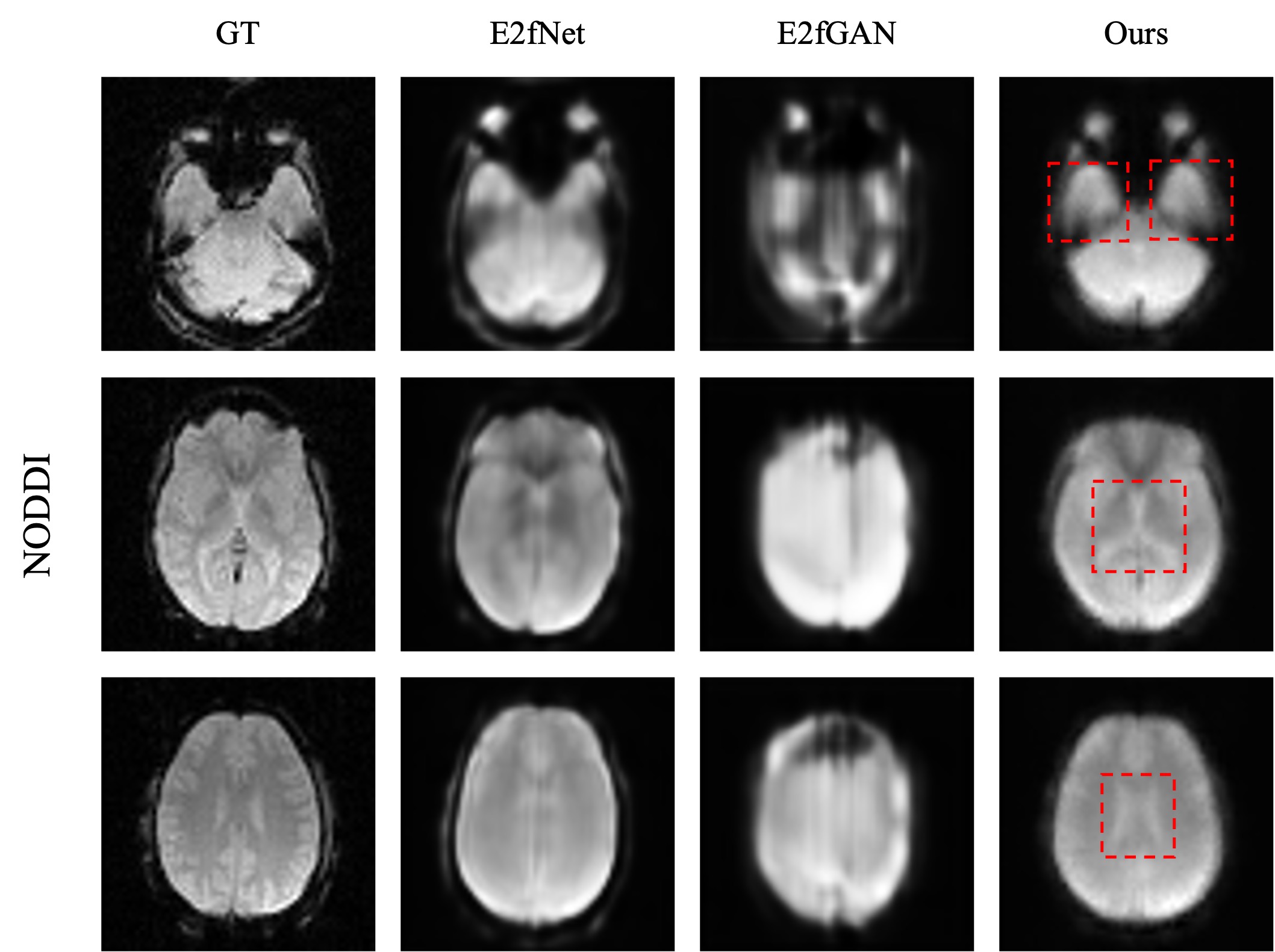}
  \caption{Visual comparison of reconstructed fMRI volumes from Spec2VolCAMU-Net and certain baseline methods on the NODDI dataset. The first column shows the ground truth (GT) fMRI volume, while the subsequent columns display the results from E2fNet, E2fGAN, and Spec2VolCAMU-Net. The brain regions that improve the most with Spec2VolCAMU-Net are highlighted with red boxes.}
  \label{fig:noddi}
\end{figure*}

\begin{figure*}[t]
  \centering
  \includegraphics[width=0.8\textwidth]{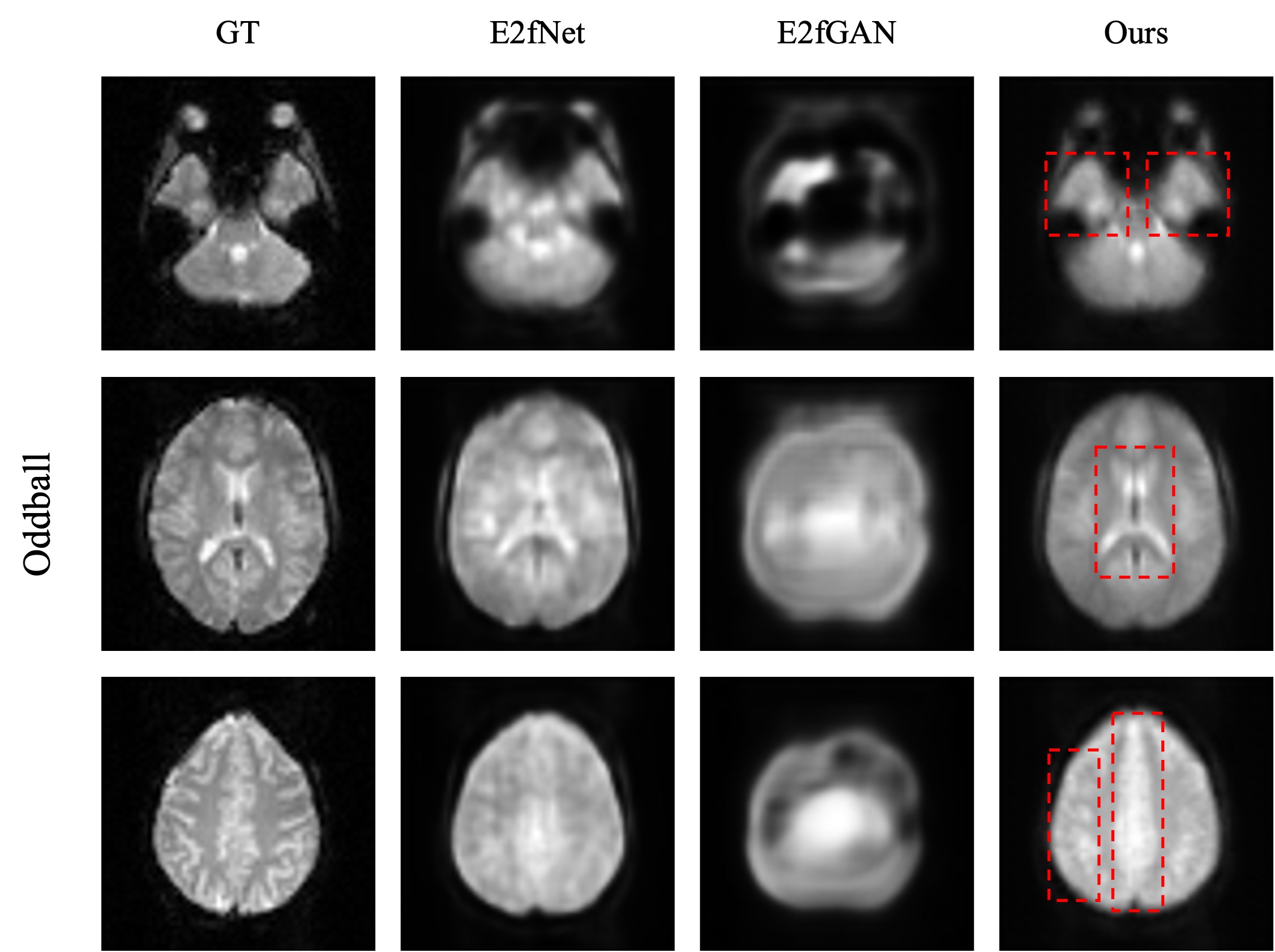}
  \caption{Visual comparison of reconstructed fMRI volumes from Spec2VolCAMU-Net and certain baseline methods on the Oddball dataset. The first column shows the ground truth (GT) fMRI volume, while the subsequent columns display the results from E2fNet, E2fGAN, and Spec2VolCAMU-Net. The brain regions that improve the most with Spec2VolCAMU-Net are highlighted with red boxes.}
  \label{fig:oddball}
\end{figure*}

To further illustrate the performance of Spec2VolCAMU-Net, we present qualitative visual comparisons of reconstructed fMRI volumes across all three datasets in Figures~\ref{fig:cn-epfl}, \ref{fig:noddi}, and \ref{fig:oddball}. The first column shows the ground truth (GT) fMRI volume, while the subsequent columns display the results from E2fNet, E2fGAN, and Spec2VolCAMU-Net. 

Specifically, on the high-resolution CN-EPFL data acquired at 3 T, as showed in Figure \ref{fig:cn-epfl}, Spec2VolCAMU-Net consistently retains both high-frequency cortical textures and deep-gray-matter anatomy that are almost indistinguishable from the ground-truth (GT) images. In the cerebellum-medulla slice, the model faithfully recovers folial subdivisions of the vermis and even the dentate nuclei, whereas E2fNet softens these details into a diffuse haze and E2fGAN collapses into a bright peripheral halo with an empty core. At the level of the lateral-ventricle bodies, Spec2VolCAMU-Net reproduces the anterior and posterior horns and the splenial notch of the corpus callosum with crisp gray-white boundaries; E2fNet rounds those contours and blurs the cortex-white-matter interface, while E2fGAN converts the central region into a feature-less luminous blob and simultaneously overexposes the periphery. In the most rostral frontal slice, Spec2VolCAMU-Net alone reinstates the fine gyral-sulcal relief: E2fNet delivers a Gaussian-like smooth surface, and E2fGAN keeps only an outline of brightness with the interior texture entirely lost.

The NODDI dataset, collected at 1.5 T and contaminated by motion-induced vertical streaks, further stresses each model's robustness. As showed in Figure \ref{fig:noddi}, Spec2VolCAMU-Net suppresses these barcode-like artifacts and still recovers both cortical folds and subcortical nuclei. In the cerebellum-pons slice, it reconstructs normal bridge-pontine contrast and temporal-lobe sulci in a noisy background, whereas E2fNet portrays the pons as an amorphous gray mass and E2fGAN drowns the entire field in striping and overexposure. In the basal-ganglia slice, the thalamic margins and the hyperintense internal capsule are sharply delineated only by Spec2VolCAMU-Net; E2fNet provides a vague outline and E2fGAN is again dominated by vertical bars. At parietal-lobe level, the central sulcus and cortical thickness appear with GT-like fidelity only in the proposed model; E2fNet remains uniformly smooth and E2fGAN preserves nothing beyond a coarse outer rim.

Oddball data, the noisiest EPI-weighted set, poses the greatest challenge. As showed in Figure \ref{fig:oddball}, Spec2VolCAMU-Net continues to avoid both over-smoothing and collapse, maintaining structural fidelity across brainstem, deep gray, and neocortex. In the cerebellar slice, the fourth ventricle and dentate nuclei are clearly identifiable, while E2fNet's rendition is dim and defocused and E2fGAN degenerates into a ring-bright cavity. At the thalamus-hippocampus level, only Spec2VolCAMU-Net exhibits distinct thalamic nuclei and a well-contoured hippocampal tail; E2fNet suffers from floating noise speckles with low contrast, and E2fGAN reduces the region to an oblong bright mass. In the occipital slice, cortical folding patterns adhere closely to GT under Spec2VolCAMU-Net, but E2fNet produces generalized blur and E2fGAN again forfeits internal texture.

Across all three datasets, common trends emerge. E2fNet invariably blurs gyral-sulcal relief and attenuates gray-white contrast, creating a soft-focus appearance. E2fGAN, though potentially crisp, is highly unstable on limited medical data: it often exhibits haloing, uniform bright blobs, or vertical stripe artifacts symptomatic of mode collapse. Spec2VolCAMU-Net uniquely reconciles sharpness with stability, preserving cortical grooves, deep-gray structures, and infratentorial detail while almost eliminating pathological artifacts; consequently, its reconstructions cluster tightly around GT with few outliers.

\subsection{Results of Ablation Study}
To dissect the contribution of each key component within the Spec2VolCAMU-Net architecture, we conducted a comprehensive ablation study on the CN-EPFL dataset. The results, detailed in Table \ref{tab:ablation}, systematically quantify the impact of the VM-UNet decoder, the MHSA module, and the individual convolutional paths within the MD-TF-CAE.

\textbf{Impact of the Vision-Mamba U-Net Decoder:} The most significant performance degradation was observed when the VM-UNet decoder was replaced with a standard U-Net architecture, composed of conventional convolutional and up-sampling layers. This change resulted in a substantial drop in SSIM from 0.788 to 0.741 and a decrease in PSNR from 23.837 to 23.112. This finding strongly validates the critical role of the VM-UNet in reconstructing high-fidelity fMRI volumes. A standard U-Net, while effective at capturing hierarchical features through its skip connections, is constrained by the local receptive field of its convolutional kernels. It struggles to explicitly model the long-range spatial dependencies that characterize distributed brain networks. In contrast, the VSS blocks within the VM-UNet, with their underlying state-space mechanism and 2D selective scan, can efficiently capture global contextual information and model complex, non-local interactions across the entire 3D volume. This capability is paramount for generating anatomically coherent and structurally accurate brain maps, as reflected by the large performance gap.

\textbf{Impact of the Multi-Head Self-Attention in the Encoder:} Removing the MHSA module from the MD-TF-CAE led to the second-largest performance drop, with SSIM falling to 0.755 and PSNR to 23.345. This underscores the importance of capturing global relationships within the EEG spectrogram before decoding. While the multi-directional convolutions excel at extracting local time-frequency patterns, the MHSA module integrates this information by attending to relationships across all channels, time steps, and frequency bins. This allows the encoder to form a holistic representation that captures how activity in one electrode/frequency band relates to another, which is neurophysiologically crucial for inferring distributed fMRI activation. Without MHSA, the model relies solely on the convolutional receptive fields to aggregate information, leading to a less comprehensive encoding and subsequently poorer reconstruction quality.

\textbf{Impact of Multi-directional Convolutional Paths:} Ablating the individual convolutional paths within the encoder—temporal, frequency, and joint time-frequency—resulted in more modest but still significant performance decreases. Removing the frequency convolution path caused the most notable drop among the three (SSIM to 0.765), suggesting that accurately modeling spectral patterns is particularly vital for the EEG-fMRI mapping. The degradation from removing the temporal (SSIM to 0.769) and joint paths (SSIM to 0.772) confirms their utility in capturing dynamic signal changes and their interactions, respectively. These results collectively demonstrate the benefit of the multi-path design. By processing the spectrogram from different perspectives in parallel, the encoder builds a more robust and comprehensive feature set than a single, monolithic convolutional block could. The fact that the model's performance degrades with the removal of any single path confirms that they learn complementary, rather than entirely redundant, features.

In conclusion, the ablation study provides compelling empirical evidence for our architectural choices. The results highlight a clear hierarchy of importance: the long-range spatial modeling of the VM-UNet decoder is most critical, followed by the global feature integration of the MHSA in the encoder, and complemented by the rich, multi-faceted local feature extraction of the directional convolutional paths. Each component makes a synergistic contribution, and their combined effect is what enables Spec2VolCAMU-Net to achieve its state-of-the-art performance.

\begin{table*}[htbp!]
  \centering
  \caption{Ablation study results on the CN-EPFL dataset.}
  \begin{tabular}{lccccccc}
    \toprule
   \multicolumn{4}{c}{\textbf{MD-TF-CAE Encoder}} & \multirow{2}{*}{\textbf{VM-UNet Decoder}} & \multirow{2}{*}{\textbf{SSIM}} & \multirow{2}{*}{\textbf{PSNR}} \\
    \cmidrule(lr){1-4}
   Temporal Conv. & Frequency Conv. & Joint Time-Freq. Conv. & MHSA & & \\
    \midrule
  \checkmark & \checkmark & \checkmark & \checkmark & \checkmark & \textcolor{red}{\textbf{0.788}} & \textcolor{red}{\textbf{23.837}} \\
  \checkmark & \checkmark & \checkmark & \checkmark & $\times$ & 0.766 & 23.185 \\
  \checkmark & \checkmark & \checkmark & $\times$ & \checkmark & 0.778 & 23.665 \\
  \checkmark & \checkmark & $\times$ & \checkmark & \checkmark & 0.787 & 23.801 \\
  \checkmark & $\times$ & $\times$ & \checkmark & \checkmark & 0.787 & 23.825 \\
  \checkmark & $\times$ & \checkmark & \checkmark & \checkmark & 0.786 & 23.481 \\
  $\times$ & $\times$ & \checkmark & \checkmark & \checkmark & 0.786 & 23.490 \\
  $\times$ & \checkmark & \checkmark & \checkmark & \checkmark & 0.769 & 23.124 \\
  $\times$ & \checkmark & $\times$ & \checkmark & \checkmark & 0.775 & 23.461\\
    \bottomrule                           
\end{tabular}
  \label{tab:ablation}
  \footnotesize{The best results are highlighted in \textcolor{red}{\textbf{red}}.}
\end{table*}

\subsection{Reasons for the Superior Performance}
The superior performance of Spec2VolCAMU-Net can be attributed to several innovative architectural designs and training strategies. Firstly, the proposed MD-TF-CAE uniquely combines lightweight multi-directional convolutions (temporal, spectral, and joint time-frequency) with multi-head self-attention. This hybrid approach effectively captures both local, fine-grained features from EEG spectrograms and global, long-range dependencies across electrode signals, leading to a more comprehensive and informative latent representation for fMRI synthesis than methodologies relying solely on CNNs or graph-attention mechanisms. Secondly, the integration of a Vision-Mamba U-Net (VM-UNet) \cite{ruan2024vm} as the decoder represents a significant step forward. By employing selective-scan state-space blocks (VSS Blocks) \cite{liu2024vmamba}, the VM-UNet architecture efficiently models spatial long-range dependencies with linear computational complexity \cite{liu2024vmamba}. This not only reduces the memory footprint, allowing for deeper and more complex decoding pathways on standard GPU hardware \cite{gu2023mamba}, but also enhances gradient flow and convergence stability due to the absence of adversarial training components \cite{mescheder2018training}. Thirdly, the utilization of a joint SSIM-MSE loss function provides multi-scale supervision during training. The SSIM component guides the model to maintain macro-structural integrity and perceptual quality, while the MSE component refines voxel-level accuracy and detail sharpness. This synergistic loss formulation encourages reconstructions that are not only quantitatively accurate but also perceptually realistic and neuro-anatomically plausible.

\subsection{Limitations and Future Work}
Despite these significant advancements, certain limitations warrant discussion and offer avenues for future research. While PSNR values are competitive, the absolute figures (20-24 dB) suggest that there is still potential for enhancing the restoration of very high-frequency details in the reconstructed fMRI volumes. The model's performance, though generally robust, showed sensitivity to severe EEG artifacts in one subject, as evidenced by an outlier in the NODDI dataset results. This highlights the need for future work on integrating more sophisticated, perhaps self-supervised, noise modeling or adaptive artifact rejection techniques directly within the reconstruction pipeline to further improve resilience to extreme cases and varying data quality. Furthermore, the current validation has predominantly focused on resting-state and relatively simple task paradigms. Extending the evaluation to more complex cognitive tasks and diverse patient populations will be crucial for rigorously assessing the generalization capabilities and broader applicability of Spec2VolCAMU-Net. 

A key limitation is that the current model and its evaluation focus on the reconstruction of individual fMRI volumes at discrete time points. Consequently, the employed metrics, SSIM and PSNR, primarily assess spatial fidelity rather than the temporal coherence of the BOLD signal. To enhance the model's relevance and persuasiveness for the neuroimaging community, particularly for applications like functional connectivity analysis, a critical future direction is to validate the temporal dynamics of the reconstructed fMRI signal. We plan to address this by extending the framework to generate entire BOLD time series. The quality of these series will be evaluated by comparing them against ground truth data using metrics such as the Pearson correlation coefficient of time courses extracted from predefined Regions of Interest (ROIs) or by computing pseudo R-squared values. Demonstrating that the reconstructed voxel time series maintain high temporal consistency and faithfully capture the underlying neural dynamics is a crucial next step that we will emphasize in our future work.

Future investigations could also explore advanced techniques such as frequency-domain self-distillation to better capture subtle spectral dynamics, or multimodal consistency regularization \cite{guan2021multimodal} to ensure tighter coupling between the latent representations of EEG and fMRI. Large-scale multi-task validation, potentially incorporating generative adversarial components in a more stable framework or exploring diffusion models \cite{croitoru2023diffusion,yang2023diffusion}, could further consolidate the practical utility and push the boundaries of EEG-to-fMRI reconstruction, paving the way for its wider adoption as a cost-effective and accessible neuroimaging tool.

\section{Conclusion}
\label{sec:conclusion}
In this paper, we presented Spec2VolCAMU-Net, a novel deep learning architecture that significantly advances the state-of-the-art in synthesizing high-fidelity 3D fMRI volumes from EEG spectrograms. The model's strength lies in its synergistic design, featuring a Multi-directional Time-Frequency Convolutional Attention Encoder (MD-TF-CAE) to capture a rich hierarchy of neuroelectric features and a Vision-Mamba U-Net decoder to efficiently model the long-range spatial dependencies inherent in brain activation maps. Our comprehensive evaluations across three diverse public datasets confirmed its superiority, establishing new benchmarks in structural similarity (SSIM) while maintaining competitive pixel-level accuracy (PSNR).

Crucially, we position the success in high-quality single-volume reconstruction as a foundational and indispensable first step. The ultimate ambition of EEG-to-fMRI synthesis is to generate temporally coherent fMRI time series suitable for dynamic brain analysis. We argue that ensuring the high spatial fidelity of each individual volumetric snapshot is a prerequisite for faithfully modeling the temporal evolution of neural activity. Therefore, this work establishes the robust spatial groundwork necessary to tackle the subsequent, more complex challenge of achieving full and accurate spatio-temporal reconstruction.

Building upon this validated spatial foundation, our future research will prioritize extending the model to generate and validate complete BOLD time series, with a focus on their temporal consistency. By continuing to refine this technology, Spec2VolCAMU-Net and its successors hold the promise of bridging the gap between high-temporal-resolution EEG and high-spatial-resolution fMRI, ultimately paving the way for more accessible, cost-effective, and powerful functional neuroimaging tools for both clinical and research communities.

\section*{Acknowledgments}
This research was funded by 
the Scientific and Technological Research Program of the Chongqing Education Commission (KJZD-K202303103, KJZD-K202501107, KJQN202501104), 
the Natural Science Foundation of Chongqing (CSTB2024NSCQ-MSX0118), 
the Chongqing Municipal Key Project for Technology Innovation and Application Development (CSTB2024TIAD-KPX0042, CSTB2025TIAD-KPX0002),
the College Student Innovation and Entrepreneurship Project (2025CXXL025), 
and the National Natural Science Foundation of P.R. China (61173184).

\section*{Declarations}\label{sec:Declarations}
\begin{itemize}
    \item \textbf{Competing Interests} All the authors declare that they have no conflict of interest.
    \item \textbf{Authors contribution statement} Conceptualization, Methodology, Writing-original draft: [Dongyi He]; Visualization, Validation, Writing-review \& editing: [Shiyang Li]; Supervision, Funding acquisition: [Bin Jiang, and He Yan].
    \item \textbf{Data Availability and Access} The datasets used in this study are publicly available. The code is available at \url{https://github.com/hdy6438/Spec2VolCAMU-Net}.
\end{itemize}

\bibliographystyle{elsarticle-num} 
\bibliography{ref}

\end{document}